\documentclass[final,3p]{elsarticle}
\usepackage{graphicx}
\usepackage{amsfonts,latexsym,eucal,amsthm,amssymb} 
\usepackage{psfrag}
\usepackage{color}
\usepackage{amsbsy}     
\usepackage{lineno}
\usepackage{flushend}
\usepackage{cuted}
\usepackage{dsfont}
\usepackage{comment}
\usepackage{mathtools}
\usepackage{natbib}

\def\dd{\mbox{d}}

\def\ve{\varepsilon}

\def\k{\kappa}

\def\NS{N_{\rm S}}
\def\NI{N_{\rm I}}
\def\Xs{N_{\rm s}}
\def\Xi{N_{\rm i}}
\def\Xsi{N_{\rm si}}
\def\Xss{N_{\rm ss}}
\def\Xii{N_{\rm ii}}
\def\rhos{\rho_{\rm s}}
\def\rhoi{\rho_{\rm i}}
\def\rhosi{\rho_{\rm si}}
\def\rhoss{\rho_{\rm ss}}
\def\rhoii{\rho_{\rm ii}}
\begin{document}
\begin{frontmatter}

 \title{Uniformly accurate effective equations for disease
    transmission mediated by pair formation dynamics}


\author{Jonathan Wylie$^{1}$ and Tom Chou$^{2}$}
\address{$^{1}$Department of Mathematics, City University of Hong
  Kong, Tat Chee Avenue, Hong Kong \\$^{2}$Dept. of Biomathematics,
  UCLA, Los Angeles, CA 90095-1766}


\begin{abstract}
  We derive and asymptotically analyze mass-action models for disease
  spread that include transient pair formation and
  dissociation. Populations of unpaired susceptibles and infecteds are
  distinguished from the population of three types of pairs of
  individuals; both susceptible, one susceptible and one infected, and
  both infected. Disease transmission can occur only within a pair
  consisting of one susceptible individual and one infected
  individual.  By considering the fast pair formation and fast pair
  dissociation limits, we use a perturbation expansion to formally
  derive a uniformly valid approximation for the dynamics of the total
  infected and susceptible populations. Under different parameter
  regimes, we derive uniformly valid effective equations for the total
  infected population and compare their results to those of the full
  mass-action model. Our results are derived from the fundamental
  mass-action system without implicitly imposing transmission
  mechanisms such as that used in frequency-dependent models.  They
  provide a new formulation for effective pairing models and are
  compared with previous models.
\end{abstract}


\end{frontmatter}




\section{Introduction}

Ordinary differential equations have been widely used to model
population biology and disease spread in systems where the agents are
spatially homogeneous. Canonical mass-action theories include the SIS,
SIR, SEIR/SEIS, and other models, which have been widely used to provide
insight into the dynamics of infected populations
\citep{MURRAY,ANDERSON1982}. Such models are simplified, averaged
representations of disease spread within complex, multispecies and
heterogeneous populations. In this paper, we revisit and analyze
transmission models and consider the effects of pairing dynamics on
infectious disease propagation through a population.

Typically, the transmission rate is assumed to depend on three
factors: a) rate at which an infected individual contacts other
individuals; b) proportion of the contacts that are with susceptible
individuals and c) the probability that a contact between the infected
individual and a susceptible individual leads to the susceptible
becoming infected. An important factor in determining the contact rate
is the relative timescales of the time required for an infected
individual to ``find'' another individual and the time required for
behavior that is responsible for transmission.

Two widely used models dominate the literature \citep{ANDERSON1982}.
\textit{Mass-action} transmission models assume that the contact rate
between any one infected individual and susceptible individuals is
proportional to the \textit{density} of susceptible individuals.  That
is, the transmission rate is given by $B_{\rm m} \rho_{\rm S}
\rho_{\rm I}$, where $\rho_{\rm S}$ is the density of susceptible
individuals, $\rho_{\rm I}$ is the density of infected individuals,
and $B_{\rm m}$ is a constant rate $\times$ area.  Mass-action models
are appropriate when the time required to ``find'' other individuals
is significantly longer than the time required for the behavior that
spreads the disease and is probably more representative of diseases
like tuberculosis (TB) where the required interaction can be just a
fleeting contact \citep{TB}.

\textit{Frequency-based} transmission models \citep{JLS2004} are
independent of density and assume that an infected individual
experiences the same number of contacts in a given time period
regardless of the density.  Given that the populations are
homogeneous, and that the contact mechanism does not distinguish
between susceptible and infected individuals, the proportion of
contacts that are with susceptibles will be $\rho_{\rm S}/(\rho_{\rm
  S}+\rho_{\rm I})$.  The transmission rate will therefore be $B_{\rm
  f}\rho_{\rm I}\rho_{\rm S}/(\rho_{\rm S}+\rho_{\rm I})$, where
$B_{\rm f}$ is a constant rate.  Frequency-based models are appropriate
when behavior that leads to disease transmission is the rate-limiting
step and is appropriate for sexually transmitted diseases in
populations that afford large numbers of partners. In other words, the
frequency of new contacts is limited by some other social/behavioral
process and not simply proportional to density. Moreover, there are
examples of populations that follow neither mass-action nor
frequency-dependent dynamics \citep{ELKS2013}, or in which contact rates
scale with density in nontrivial ways \citep{HU2013}.


Various authors have also proposed models with infection rates of the
form $B_{\rm f} B_{\rm m}\rho_{\rm I}\rho_{\rm S}/\left[B_{\rm
    f}+B_{\rm m}(\rho_{\rm S}+\rho_{\rm I})\right]$.  Such terms are
similar to Holling's Type II functional response in predator-prey
models and asymptotically reproduce the mass-action transmission for
low densities and the frequency-based transmission for large densities
\citep{DAWES2013}.  Whether or not this type of Holling's type II model
can be theoretically justified or whether an alternative functional
response is more natural is also an important theoretical question.
Heuristic frequency-based and Holling-type models implicitly incorporate
behavior into the dynamics. However, qualitatively, one expects that these
dynamics might arise from higher-dimensional mass-action ODEs that
explicitly include intermediate ``reactions'' that reflect some of
behavioral processes. 

Here, we ask how can such behavior-induced frequency-based models be
formally derived from the fundamental mass-action process by
considering the simplest mass-action model in which {\it lone}
susceptible and infected individuals associate to form pairs (either
susceptible-susceptible, susceptible-infected or infected-infected)
\citep{PAIRMODEL1988,HEESTERBEEK1993,PAIRMODEL0,GHANI2010}.  These
models are similar to the class of household structure models in which
groups of individuals form subgroups within which disease transmission
spreads faster
\citep{SANDER2004,HOUSEHOLD2007,FRASER2007,HOUSEHOLD2008}. Pairs can
also dissociate into their constituent lone individuals. In these pair
formation type models, transmission can only occur from an infected to
a susceptible in a susceptible-infected pair. We also include the
effects of death and immigration of susceptibles which leads directly
to a set of five differential equations: two ODEs for two types of
lone individuals and three ODEs for the three different types of
pairs.  How these five equations can be reduced to \textit{effective}
equations under certain conditions will be the topic of our analyses.
Previous treatments of the pairing models have been put forth but
either do not consider certain parameter limits
\citep{HEESTERBEEK1993}, are not systematic \citep{JLS2004}, implicitly
force a frequency-dependent interaction through a ``mixing matrix''
\citep{JLS2004}, or only provide approximations at short times
\citep{EARLY2014}.


In this paper, we generate effective, mass-action-derived equations
that are uniformly valid at all times.  We first show that if pair
dissociation and within-pair transmission is fast, to lowest order,
the equations simply reduce to two mass-action-like equations, one for
the total infected density and one for the total susceptible density,
but with an effective transmission coefficient.  Although no
structural change is seen in this case, if association and
dissociation are asymptotically faster than the other processes
(death, transmission, and immigration), the resulting effective
equations for the total infected and susceptible populations involve
terms of rational fractions of polynomials. These equations further
reduce to simpler forms in certain parameter limits.

On the other hand, if association is asymptotically faster than the
other process (including dissociation), we show that the leading-order
dynamics can only be reduced to three ODEs that bear a number of
similarities to models that include an \textit{exposed} subpopulation,
such as the SEI (Susceptible-Exposed-Infected) class of models. This
new type of model, derived from the fundamental mass-action pairing
model, reflects a latency period in disease propagation but is still
different from the typical SEI-type model.

%
%
%

\section{Models}

We start by reviewing the basic mass-action, frequency-based, and
pairing models for disease propagation.

\subsection{Simple Mass-Action Model}
The simplest mass-action description for the dynamics of the
susceptible and infected population densities $\rhos(t)$ and
$\rhoi(t)$ is given by the susceptible-infected (SI) model with
immigration:

\begin{align}
\displaystyle  {\dd \rhos(t) \over \dd t}  &= \tilde{\Pi} -\mu_{\rm s} \rhos(t)  -
B_{\rm m} \rhos(t) \rhoi(t) \nonumber \\
\displaystyle  {\dd \rhoi(t) \over \dd t}   &=  -\mu_{\rm i} \rhoi(t)  +
B_{\rm m} \rhos(t) \rhoi(t),
\label{MassAction}
\end{align}
where $\tilde{\Pi}$ represents the rate at which the density of
susceptible individuals increases via immigration from outside the
region, and $\mu_{\rm s}$ and $\mu_{\rm i}$ are the death rates of
susceptible and infected individuals, respectively.

If recovery of infecteds back to the susceptible pool is included,
Eqs.~\ref{MassAction} becomes the standard SIS model when $\tilde{\Pi}
= \mu_{\rm s}=\mu_{\rm i} = 0$ and the total population is conserved.
The steady-state solution to Eqs.~\ref{MassAction},
$(\rhos^{*},\rhoi^{*})=(\tilde{\Pi}/\mu_{\rm s},0)$, exists for all
parameters and is linearly stable if the reproduction number

\begin{equation}
R_{\rm m} =\frac{B_{\rm m}\tilde{\Pi}}{\mu_{\rm s}\mu_{\rm i}}<1
\end{equation}
and linearly unstable if $R_{\rm m}>1$.  A second steady state
$(\rhos^{*},\rhoi^{*})=(\mu_{\rm i}/B_{\rm m},\tilde{\Pi}/\mu_{\rm
  i}-\mu_{\rm s}/B_{\rm m})$ only has positive densities and hence
exists for $R_{\rm m}>1$ and is linearly stable.  For values of
$R_{\rm m}>1$, a non-zero infected population can be maintained
indefinitely, whereas for $R_{\rm m}<1$, the infected population will
ultimately die out.

\subsection{Frequency-Dependent Model}
A typical frequency-based model takes the form
\begin{align}
\begin{split}
\displaystyle  {\dd \rhos\over \dd t}  & = \tilde{\Pi} -\mu_{\rm s}\rhos  -
B_{\rm f}\frac{\rhos\rhoi}{\rhos+\rhoi} \\
\displaystyle  {\dd \rhoi \over \dd t}   & =-\mu_{\rm i}\rhoi  +
B_{\rm f}\frac{\rhos\rhoi}{\rhos+\rhoi},
\label{FixedFreq}
\end{split}
\end{align}
which is often used to describe sexually transmitted disease in which
the pair formation rate is thought to be intrinsic to the individual
and largely population density-independent.

The steady state $(\rhos^{*},\rhoi^{*})=(\Pi/\mu_{\rm s},0)$ exists
for all parameters and is linearly stable if
\begin{equation}
R_{\rm f}\coloneqq \frac{B_{\rm f}}{\mu_{\rm i}}<1
\end{equation}
and linearly unstable if $R_{\rm f}>1$. A second steady state

\begin{equation}
(\rhos^{*},\rhoi^{*}) = \left(\frac{\tilde{\Pi}}{\mu_{\rm s}+B_{\rm f}-\mu_{\rm
      i}},\frac{(B_{\rm f}-\mu_{\rm i})\tilde{\Pi}} {\mu_{\rm i}(\mu_{\rm
      s}+B_{\rm f}-\mu_{\rm i})}\right)
\end{equation}
has only positive densities and hence exists for $R_{\rm f}>1$ and is
linearly stable.

An important difference between the mass-action and frequency-based
models is hence apparent. Under mass-action, the reproduction number
depends on the influx of individuals $\Pi$ and so reducing the influx
of individuals will be an effective strategy in disease control. On
the other hand, for the frequency model, the reproduction number is
independent on the influx and so reducing the influx will not cause
the disease to die out.

\subsection{Mass-Action Pairing Model}
We now consider the simplest mass-action model that explicitly
includes population \textit{densities} of transient pairs:

%

\begin{align}
\begin{split}
\displaystyle {\dd \rhos \over \dd t} &= \tilde{\Pi} -\mu_{\rm s}\rhos -
2\tilde{a}_{\rm ss}\rhos^2 - \tilde{a}_{\rm si}
\rhos\rhoi + 2(\mu_{\rm ss}+d_{\rm ss})\rhoss +
(\mu_{\rm is}+d_{\rm si})\rhosi \\
\displaystyle {\dd \rhoi \over \dd t} &= -\mu_{\rm i}\rhoi
-2\tilde{a}_{\rm ii}\rhoi^2 -
\tilde{a}_{\rm si}\rhos\rhoi + 2(\mu_{\rm ii}+d_{\rm ii})\rhoii +
(\mu_{\rm si}+d_{\rm si})\rhosi \\
\displaystyle {\dd \rhoss \over \dd t} & =
-(2\mu_{\rm ss}+d_{\rm ss})\rhoss + \tilde{a}_{\rm ss}\rhos^2 \\
\displaystyle {\dd \rhosi \over \dd t} &= -(\mu_{\rm is}+\mu_{\rm si} +
d_{\rm si}+\beta)\rhosi + \tilde{a}_{\rm si}\rhos\rhoi \\
\displaystyle {\dd \rhoii \over \dd t} &= -(2\mu_{\rm ii} +
d_{\rm ii})\rhoii + \beta \rhosi + \tilde{a}_{\rm ii}\rhoi^2,
\label{EQNRHO}
\end{split}
\end{align}
where $\rhos$ and $\rhoi$ are the densities of lone susceptible and
infected individuals respectively. The quantities $\rhoss$, $\rhosi$
and $\rhoii$ are the densities of susceptible-susceptible,
susceptible-infected and infected-infected pairs, respectively.  In
this model, transmission can occur only from infecteds to susceptibles
who are in a susceptible-infected pair and happens at rate $\beta$.
The rate of immigration of density of lone susceptibles is denoted by
$\tilde{\Pi}$. In Eqs.~\ref{EQNRHO}, $\mu$ represent the death rates
of the indicated species; for example, $\mu_{\rm si}$ is the death
rate for a susceptible in a susceptible-infected pair, and $\mu_{\rm
  is}$ is the death rate for an infected in a susceptible-infected
pair.
%
%
The quantities $d$ represent the dissociation rate of the indicated
pairs while $\tilde{a}$, because they denote an interaction between
two individuals and multiply terms quadratic in density, are
association rates \textit{per density} and have units of rate $\times$
area.


In order to analyze the full model, we nondimensionalize by
multiplying each equation by a reference area $A_{0}$:


\begin{subequations}
\begin{align}
\displaystyle {\dd \Xs \over \dd t} &= \Pi -\mu_{\rm s}\Xs -
2 a_{\rm ss}\Xs^2 - a_{\rm si}
\Xs\Xi + 2(\mu_{\rm ss}+d_{\rm ss})\Xss +
(\mu_{\rm is}+d_{\rm si})\Xsi \label{EQNNa}\\
\displaystyle {\dd \Xi \over \dd t} &= -\mu_{\rm i}\Xi
-2a_{\rm ii}\Xi^2 -a_{\rm si}\Xs\Xi 
+ 2(\mu_{\rm ii}+d_{\rm ii})\Xii +
(\mu_{\rm si}+d_{\rm si})\Xsi \label{EQNNb} \\
\displaystyle {\dd \Xss \over \dd t} & =
-(2\mu_{\rm ss}+d_{\rm ss})\Xss + a_{\rm ss}\Xs^2 \label{EQNNc} \\
\displaystyle {\dd \Xsi \over \dd t} &= -(\mu_{\rm is}+\mu_{\rm si} +
d_{\rm si}+\beta)\Xsi + a_{\rm si}\Xs\Xi \label{EQNNd} \\
\displaystyle {\dd \Xii \over \dd t} &= -(2\mu_{\rm ii} +
d_{\rm ii})\Xii + \beta \Xsi + a_{\rm ii}\Xi^2,
\label{EQNNe}
\end{align}
\end{subequations}
where $N = \rho A_{0}, a \equiv \tilde{a}/A_{0}$, and $\Pi =
A_{0}\tilde{\Pi}$ are the dimensionless populations within area
$A_{0}$, the rate of association (with units of 1/time), and the
immigration rate per density, respectively. The reference area $A_{0}$
is arbitrary, but can be chosen to scale the magnitudes of $N$ and the
relative rates $a/\mu$.  Under any particular scaling,
different limits of the magnitudes of $N$ and $a/\mu, d/\mu$ can be
used to further analyze Eqs.~\ref{EQNNa}-\ref{EQNNe}.

Note that if one mixes the mass-action model with frequency-dependent
transmission, as has been often done \citep{JLS2004,PAIRMODEL0}, the
quadratic pairing terms in Eqs.~\ref{EQNRHO} would be replaced by,
\textit{e.g.}, $\tilde{a}_{\rm si}\rhoi\rhos/(\rhoi+\rhos) =
\tilde{a}_{\rm si}\Xi\Xs/(\Xi+\Xs)$, where here, $\tilde{a}_{\rm si}$
has units of 1/time.

\section{Asymptotic Analyses and Discussion}

We now analyze the mass-action pairing model in different limits to
reduce the model to simpler forms in order to illustrate how pairing
and dissociation affect the overall propagation of infection.

\subsection{Fast Dissociation and Transmission Limit}

First, consider the simplest case where the dissociation and
transmission rates are large by scaling them according to $d_{\rm ss}
= \bar{d}_{\rm ss}/\ve, d_{\rm si} = \bar{d}_{\rm si}/\ve, d_{\rm ii}
= \bar{d}_{\rm ii}/\ve$, and $\beta =\bar{\beta}/\ve$, with $\ve \to
0^{+}$. In this limit, we expect the number or density of pairs to be
much smaller than the number of unpaired individuals. We adopt
an expansion of the form

\begin{align}
\begin{split}
\Xs &= \Xs^{(0)} + \ve \Xs^{(1)}+ \cdots\\
\Xi &= \Xi^{(0)} + \ve \Xi^{(1)} +\cdots\\
\Xss &= \Xss^{(0)} + \ve \Xss^{(1)} +\cdots\\
\Xsi &= \Xsi^{(0)} + \ve \Xsi^{(1)} +\cdots\\
\Xii &= \Xii^{(0)} + \ve \Xii^{(1)}+ \cdots
\label{EQN2}
\end{split}
\end{align}
and substitute it into Eqs.~\ref{EQNNa}-\ref{EQNNe}.
To leading order, we obtain $\Xss^{(0)}= \Xsi^{(0)}=\Xii^{(0)}=0$, while to
the next order, we find

\begin{equation}
\Xss^{(1)} = {a_{\rm ss} \over \bar{d}_{\rm ss}}\Xs^{(0)2},
\quad \Xsi^{(1)} = {a_{\rm si} \over \bar{\beta}+ \bar{d}_{\rm
    si}}\Xs^{(0)}\Xi^{(0)}, \quad \Xii^{(1)} = {\bar{\beta}\over \bar{d}_{\rm
    ii}}\Xsi^{(1)} + {a_{\rm ii} \over \bar{d}_{\rm ii}}\Xi^{(0)2}.
\end{equation}

Substitution of the above approximations for the pair
populations into the scaled equations for $\Xs$ and $\Xi$ (derived
from Eqs.~\ref{EQNNa}-\ref{EQNNb}), we find to lowest order

\begin{align}
\begin{split}
\displaystyle {\dd \Xs^{(0)} \over \dd t} & = \Pi -\mu_{\rm
  s}\Xs^{(0)} -\left({a_{\rm si}\beta \over
  \beta+d_{\rm si}}\right)\Xs^{(0)}\Xi^{(0)} \\ \displaystyle {\dd \Xi^{(0)}
  \over \dd t} & = -\mu_{\rm i}\Xi^{(0)} + \left({a_{\rm
    si}\beta\over \beta+d_{\rm si}}\right)\Xs^{(0)}\Xi^{(0)}, 
\label{EQNEFF}
\end{split}
\end{align}
wherein an \textit{effective} transmission rate can be defined 
as

\begin{equation}
  B_{\rm eff} \coloneqq {a_{\rm si} \beta \over \beta + d_{\rm si}}=
       {a_{\rm si} \bar{\beta} \over \bar{\beta} + \bar{d}_{\rm si}}.
  \label{BETAEFF}
\end{equation}
In this limit, the effective equations for infecteds and susceptibles
retain the mass-action form, but with a modified transmission
parameter.  The pair formation process mediates the disease
transmission through the association rate $a_{\rm si}$. For
$\bar{\beta} \ll \bar{d}$, the rate limiting step is transmission
within a susceptible-infected pair. When intrapair transmission is
fast, $\bar{\beta} \gg \bar{d}$, the overall transmission rate $B_{\rm
  eff} \approx a_{\rm si}$ approaches the association rate itself.
Thus, in this limit, the five-dimensional mass-action pairing
equations reduce to a two-dimensional mass-action model with a
modified transmission rate.  Note that if we were to use the
frequency-dependent variant of the pairing model, the form would also
be preserved to lowest order with the corresponding transmission term
$B_{\rm eff}\Xs^{(0)}\Xi^{(0)}/(\Xi^{(0)}+\Xs^{(0)})$.

\subsection{Fast Association and Dissociation Limit}
We now consider the limit where \textit{both} the association and dissociation
coefficients are significantly larger than the death and infection
rates and define $a_{\rm ss}=\bar{a}_{\rm ss}/\ve$, $a_{\rm
  si}=\bar{a}_{\rm si}/\ve$, $a_{\rm ii}=\bar{a}_{\rm ii}/\ve$,
$d_{\rm ss}= \bar{d}_{\rm ss}/\ve$, $d_{\rm si}=\bar{d}_{\rm si}/\ve$
and $d_{\rm ii}=\bar{d}_{\rm ii}/\ve$, with $\ve \to 0^{+}$.  We also
perform a linear transformation on Eqs.~\ref{EQNNa}-\ref{EQNNb} so that they
describe \textit{total} susceptible and infected populations and are
independent of $\ve$:

\begin{align}
\begin{split}
\displaystyle  {\dd \over \dd t}\left(\Xs+2\Xss+\Xsi \right) &
= \Pi -\mu_{\rm s}\Xs - 2\mu_{\rm ss}\Xss -
(\mu_{\rm si}+\beta)\Xsi \\
\displaystyle  {\dd \over \dd t}\left(\Xi + 2\Xii+\Xsi \right)  
 &= -\mu_{\rm i}\Xi - 2\mu_{\rm ii}\Xii     -
(\mu_{\rm is}-\beta)\Xsi \\
\displaystyle  {\dd \Xss \over \dd t}   & = -\left(2\mu_{\rm ss}+
\frac{\bar{d}_{\rm ss}}{\ve}\right)\Xss +
\frac{\bar{a}_{\rm ss}}{\ve}\Xs^2 \\
\displaystyle  {\dd \Xsi  \over \dd t}   &= -\left(\mu_{\rm is}+\mu_{\rm si} 
+\frac{\bar{d}_{\rm si}}{\ve} +\beta\right)\Xsi +
\frac{\bar{a}_{\rm si}}{\ve}\Xs\Xi \\
\displaystyle  {\dd \Xii \over \dd t}   &=  -\left(2\mu_{\rm ii} + 
\frac{\bar{d}_{\rm ii}}{\ve}\right)\Xii  +\beta \Xsi  +
\frac{\bar{a}_{\rm ii}}{\ve}\Xi^2.
\label{EQN1}
\end{split}
\end{align}

We now substitute the expansion in Eqs.~\ref{EQN2} into
Eqs.~\ref{EQN1} and keep only the $O(1)$ terms to find

\begin{subequations}
\begin{align}
\displaystyle {\dd \over \dd
  t}\left(\Xs^{(0)}+2\Xss^{(0)}+\Xsi^{(0)} \right) &= \Pi
-\mu_{\rm s}\Xs^{(0)} - 2\mu_{\rm ss}\Xss^{(0)} -
(\mu_{\rm si}+\beta)\Xsi^{(0)} \label{EQN3a} \\
\displaystyle {\dd \over \dd t}\left(\Xi^{(0)} +
2\Xii^{(0)}+\Xsi^{(0)} \right) &= -\mu_{\rm i}\Xi^{(0)}
-2\mu_{\rm ii}\Xii^{(0)} -(\mu_{\rm is}-\beta)\Xsi^{(0)}\label{EQN3b} \\
0 & = -\bar{d}_{\rm ss}\Xss^{(0)} + \bar{a}_{\rm ss}\Xs^{(0)2} \label{EQN3c}\\
0 & = -\bar{d}_{\rm si}\Xsi^{(0)} + \bar{a}_{\rm si}\Xs^{(0)}\Xi^{(0)}\label{EQN3d} \\
0 & = -\bar{d}_{\rm ii}\Xii^{(0)} + \bar{a}_{\rm ii}\Xi^{(0)2}.
\label{EQN3e}
\end{align}
\end{subequations}
By using Eqs.~\ref{EQN3c}-\ref{EQN3e} to eliminate $\Xss^{(0)}$,
$\Xsi^{(0)}$ and $\Xii^{(0)}$ from Eqs.~\ref{EQN3a}-\ref{EQN3b}, we
obtain

\begin{align}
\begin{split}
\displaystyle {\dd \over \dd
  t}\left(\Xs^{(0)}+2\kappa_{\rm ss}\Xs^{(0)2}+\kappa_{\rm si}\Xs^{(0)}\Xi^{(0)}
\right)  = & \Pi -\mu_{\rm s}\Xs^{(0)} -
2\mu_{\rm ss}\kappa_{\rm ss}\Xs^{(0)2}  \\
\: & \hspace{3mm} -
(\mu_{\rm si}+\beta)\kappa_{\rm si}\Xs^{(0)}\Xi^{(0)} \\
\displaystyle {\dd \over \dd t}\left(\Xi^{(0)} +
2\kappa_{\rm ii}\Xi^{(0)2}+\kappa_{\rm si}\Xs^{(0)}\Xi^{(0)}
\right) = & -\mu_{\rm i}\Xi^{(0)} -2\mu_{\rm ii}\kappa_{\rm ii}\Xi^{(0)2} \\
\: & \hspace{3mm} -
(\mu_{\rm is}-\beta)\kappa_{\rm si}\Xs^{(0)}\Xi^{(0)},
\label{EQN4}
\end{split}
\end{align}
where $\kappa_{\rm ss}=\bar{a}_{\rm ss}/\bar{d}_{\rm ss}$,
$\kappa_{\rm si}\equiv \kappa_{\rm is}=\bar{a}_{\rm si}/\bar{d}_{\rm si}$, and  $\kappa_{\rm ii}=
\bar{a}_{\rm ii}/\bar{d}_{\rm ii}$.

\subsubsection{Steady states and stability}
The most convenient way to determine the steady states and/or further
analyze Eqs.~\ref{EQN4} is to unpack them in terms of $\Xi^{(0)}$ and
$\Xs^{(0)}$ and write them in the form

\begin{equation}
   {\dd \over \dd t}
   \left[ {\begin{array}{c}
   \Xs^{(0)}\\
   \Xi^{(0)}
   \end{array} } \right]
   =
  {\bf M}^{-1}
   \left[ {\begin{array}{c}
   \Pi -\mu_{\rm s}\Xs^{(0)} - 2\mu_{\rm ss}\kappa_{\rm ss}\Xs^{(0)2} 
- (\mu_{\rm si}+\beta)\kappa_{\rm si}\Xs^{(0)}\Xi^{(0)}\\
   -\mu_{\rm i}\Xi^{(0)}-2\mu_{\rm ii}\kappa_{\rm ii}\Xi^{(0)2}
     - (\mu_{\rm is}-\beta)\kappa_{\rm si}\Xs^{(0)}\Xi^{(0)}
     \end{array} } \right],
   \label{EQN4B}
 \end{equation}
 where
\begin{equation*}
{\bf M}= \left[ {\begin{array}{cc}
   1+4\kappa_{\rm ss}\Xs^{(0)}+\kappa_{\rm si}\Xi^{(0)} & \kappa_{\rm si}\Xs^{(0)} \\
   \kappa_{\rm si}\Xi^{(0)} & 1+4\kappa_{\rm ii}\Xi^{(0)}+\kappa_{\rm si}\Xs^{(0)}
   \end{array} } \right] 
\end{equation*}
Note that since $\Xs^{(0)}, \Xi^{(0)}, \kappa_{\rm ss}, \kappa_{\rm
  si}, \kappa_{\rm ii} \ge 0$, the eigenvalues of ${\bf M}$ can never
be zero and ${\bf M}$ is invertible.

One can readily show that the system of equations always supports an
infection-free steady-state solution:

\begin{equation}
\left(\Xs^{(0)},\Xi^{(0)}\right) = 
\left(\frac{-\mu_{\rm s}+\sqrt{\mu_{\rm s}^2+8\mu_{\rm ss}
\kappa_{\rm ss}\Pi}}{4\mu_{\rm ss}\kappa_{\rm ss}},0\right),
\end{equation}
and that this solution is linearly stable
if

\begin{equation}
R \coloneqq \frac{\kappa_{\rm si}\left(\beta-\mu_{\rm is}\right)
\left(-\mu_{\rm s}+\sqrt{\mu_{\rm s}^2+8\mu_{\rm ss}\kappa_{\rm ss}\Pi}\right)}
{4\mu_{\rm i}\mu_{\rm ss}\kappa_{\rm ss}}<1.
\end{equation}
and linearly unstable if $R>1$.  Another stable solution with positive
$\Xs^{(0)}$ and $\Xi^{(0)}$ will exists if $R>1$.  This solution
structure closely mirrors that of the mass-action and
frequency-dependent models.

\subsubsection{Comparison to mass-action and frequency-based models}

In order to compare Eqs.~\ref{EQN4} or \ref{EQN4B} to
the simpler classic models, it is preferable to rewrite the equations
in terms of the leading-order expressions for the total 
susceptible and infected populations 
\begin{align}
\begin{split}
\NS^{(0)}&= \Xs^{(0)}+2\Xss^{(0)}+\Xsi^{(0)}\\
\NI^{(0)}&= \Xi^{(0)} + 2\Xii^{(0)}+\Xsi^{(0)}.
\label{TotalDensitiesLeadingOrderDef}
\end{split}
\end{align}
Again using Eqs.~\ref{EQN3c}-\ref{EQN3e} to eliminate
$\Xss^{(0)}$, $\Xsi^{(0)}$, and $\Xii^{(0)}$, we find

\begin{subequations}
\begin{align}
  \NS^{(0)}&= \Xs^{(0)}+2\kappa_{\rm ss}\Xs^{(0)2}+\kappa_{\rm si}\Xs^{(0)}\Xi^{(0)}
  \label{TotalDensitiesLeadingOrdera} \\
\NI^{(0)}&=\Xi^{(0)} + 2\kappa_{\rm ii}\Xi^{(0)2}+\kappa_{\rm si}\Xs^{(0)}\Xi^{(0)}.
\label{TotalDensitiesLeadingOrderb}
\end{align}
\end{subequations}
Next, we need to express the quantities $\Xs^{(0)}$ and $\Xi^{(0)}$
in terms of $\NS^{(0)}$ and $\NI^{(0)}$.  Solving
Eq.~\ref{TotalDensitiesLeadingOrderb} for $\Xi^{(0)}$ and substituting
the result into Eq.~\ref{TotalDensitiesLeadingOrdera}, we find a
quartic equation for $\Xs^{(0)}$

\begin{align}
\begin{split}
2\kappa_{\rm ss}(4\kappa_{\rm ii}\kappa_{\rm ss}-\kappa_{\rm si}^2)\Xs^{(0)4} +
(8\kappa_{\rm ii}\kappa_{\rm ss}-2\kappa_{\rm si}\kappa_{\rm ss}-\kappa_{\rm si}^2)\Xs^{(0)3}\hspace{1.5cm}\\
\:\hspace{1cm} + \left(\kappa_{\rm si}^2(\NS^{(0)}-\NI^{(0)})-8\NS^{(0)}\kappa_{\rm ii}\kappa_{\rm ss}
+2\kappa_{\rm ii}-\kappa_{\rm si}\right)\Xs^{(0)2} \hspace{7mm} \\
\: \hspace{1cm} +\NS^{(0)}(\kappa_{\rm si}-4\kappa_{\rm ii})\Xs^{(0)} +
2\NS^{(0)2}\kappa_{\rm ii}=0.
\label{quartic}
\end{split}
\end{align}
One can readily show that only one of the four roots gives values of
$\Xs^{(0)}$ and $\Xi^{(0)}$ that are both positive when $\NS^{(0)}$ and $\NI^{(0)}$
are positive.  Upon using this physical root for $\Xs^{(0)}$ as
functions of $\NS^{(0)}$ and $\NI^{(0)}$ in
Eq.~\ref{TotalDensitiesLeadingOrdera}, we find the unique physical
root for $\Xi^{(0)}$, expressed in terms of $\NS^{(0)}$ and
$\NI^{(0)}$.
%
Explicit formulae for the solution of a quartic are known and so we
can express $\Xs^{(0)} \equiv F_{\rm S}(\NS^{(0)},\NI^{(0)})$ and
$\Xi^{(0)}\equiv F_{\rm I}(\NS^{(0)},\NI^{(0)})$ as functions $F_{\rm
  S}$ and $F_{\rm I}$ that are obtained by the procedure described
above.  One can then rewrite
\begin{align}
\begin{split}
\displaystyle  {\dd \NS^{(0)}\over \dd t}  &= \Pi -\mu_{\rm s}\NS^{(0)}  
+ 2(\mu_{\rm s}-\mu_{\rm ss})\kappa_{\rm ss}F_{\rm S}^2   
+ (\mu_{\rm s}-\mu_{\rm si}-\beta)\kappa_{\rm si}F_{\rm S} F_{\rm I} \\
\displaystyle  {\dd \NI^{(0)} \over \dd t}   &= -\mu_{\rm i}N_{\rm I}^{(0)}
 +2(\mu_{\rm i} - \mu_{\rm ii})\kappa_{\rm ii} F_{\rm I}^2  + (\mu_{\rm i} 
-\mu_{\rm is}+\beta)\kappa_{\rm si}F_{\rm S} F_{\rm I}.
\label{EQN5}
\end{split}
\end{align}
Although $F_{\rm S}(\NS^{(0)},\NI^{(0)})$ and $F_{\rm
  I}(\NS^{(0)},\NI^{(0)})$ are unwieldy functions of $\NS^{(0)}$ and
$\NI^{(0)}$, Eqs.~\ref{EQN5} represent a systematic projection of the
original five-dimensional problem to two equations describing the
total susceptible and infected populations. These two equations can be
further simplified in the following limits.

\subsubsection{Low density asymptotics}
Consider the solutions to $\Xs^{(0)}$ and $\Xi^{(0)}$ in the limit
where the populations in the reference area $A_{0}$ are small,
$\NS^{(0)}, \NI^{(0)} \ll 1$. Upon Taylor expansion of the solutions
to Eqs.~\ref{TotalDensitiesLeadingOrdera} and
\ref{TotalDensitiesLeadingOrderb}, we find $F_{\rm
  S}(\NS^{(0)},\NI^{(0)})\approx \NS^{(0)}-(\kappa_{\rm
  si}\NI^{(0)}+2\kappa_{\rm ss}\NS^{(0)})\NS^{(0)} +O(N_{\rm
  S,I}^{(0)3})$ and $F_{\rm I}(\NS^{(0)},\NI^{(0)})\approx \NI^{(0)}
-(\kappa_{\rm si}\NS^{(0)}+2\kappa_{\rm ii}\NI^{(0)})\NI^{(0)}
+O(N_{\rm S,I}^{(0)3})$, and Eqs.~\ref{EQN5} to lowest order becomes

\begin{align}
\begin{split}
\displaystyle  {\dd \NS^{(0)}\over \dd t}  &\approx \Pi -\mu_{\rm s}\NS^{(0)}
+ 2(\mu_{\rm s}-\mu_{\rm ss})\kappa_{\rm ss}\NS^{(0)2}   +
(\mu_{\rm s}-\mu_{\rm si}-\beta)\kappa_{\rm si}\NS^{(0)}\NI^{(0)} \\
\displaystyle  {\dd \NI^{(0)} \over \dd t}   &\approx -\mu_{\rm i}\NI^{(0)}
+2(\mu_{\rm i} - \mu_{\rm ii})\kappa_{\rm ii}\NI^{(0)2}  +
(\mu_{\rm i} -\mu_{\rm is}+\beta)\kappa_{\rm si}\NS^{(0)}\NI^{(0)}
\label{SISSmallLimit}
\end{split}
\end{align}
The dynamics in this low-density limit are dominated by immigration
and death, but are also qualitatively different from those of the
standard mass-action model in that Eqs.~\ref{SISSmallLimit} contain
$N_{\rm S}^{(0)2}$ and $N_{\rm I}^{(0)2}$ terms.  These quadratic
terms arise from the difference in death rates between paired and
unpaired susceptible individuals $\mu_{\rm s}-\mu_{\rm ss}$ and paired
and unpaired infected individuals $\mu_{\rm i}-\mu_{\rm ii}$. However,
if we assume that the death rate is independent of the pairing status,
\textit{i.e.}, $\mu_{\rm ss}=\mu_{\rm s}$, $\mu_{\rm ii}=\mu_{\rm i}$,
$\mu_{\rm si}=\mu_{\rm s}$ and $\mu_{\rm is}=\mu_{\rm i}$, we obtain
the standard mass-action model with $B_{\rm m} \propto \kappa_{\rm si}\beta$.

\subsubsection{High density asymptotics}

If $\NS^{(0)}, \NI^{(0)}\gg 1$, and hence $\Xs^{(0)}, \Xi^{(0)} \gg
1$, the physical solutions to Eqs.~\ref{TotalDensitiesLeadingOrdera}
and \ref{TotalDensitiesLeadingOrderb} are approximately

\begin{align}
\begin{split}
  F_{\rm S}(\NS^{(0)},\NI^{(0)})
  & \approx \sqrt{\frac{\NI^{(0)}+(2K-1)\NS^{(0)} -
    \sqrt{\left(\NI^{(0)}-\NS^{(0)}\right)^2+4K\NS^{(0)}\NI^{(0)}}}
  {4\kappa_{\rm ss}(K-1)}}, \\
F_{\rm I}(\NS^{(0)},\NI^{(0)}) & \approx \sqrt{\frac{
    \NS^{(0)}+(2K-1)\NI^{(0)} -
    \sqrt{\left(\NI^{(0)}-\NS^{(0)}\right)^2+4K \NS^{(0)}\NI^{(0)}}}{4\kappa_{\rm ii}(K-1)}}.
\label{SILargeLimit}
\end{split}
\end{align}
%
%
where $K\equiv 4\kappa_{\rm ss}\kappa_{\rm ii}/\kappa_{\rm si}^2$ Upon
substituting Eqs.~\ref{SILargeLimit} into Eqs.~\ref{EQN5}, we find the
effective, though unwieldy, equations for $\NS^{(0)}, \NI^{(0)}\gg 1$.
  %
%
  In this case, even if $\mu_{\rm ss}=\mu_{\rm s}$, $\mu_{\rm
    ii}=\mu_{\rm i}$, $\mu_{\rm si}=\mu_{\rm s}$ and $\mu_{\rm
    is}=\mu_{\rm i}$, the effective model differs significantly in
  form from both the mass-action and frequency-dependent models.
\begin{figure*}[htb]
\begin{center}
\includegraphics[width=4.4in]{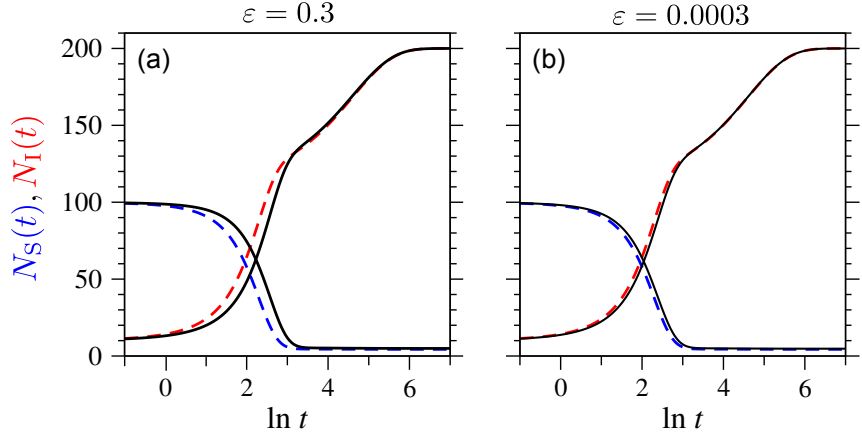}
\caption{{\bf Fast association and aissociation - high density limit:}
  Comparison of the exact numerical solution of Eqs.~\ref{EQNNa}-\ref{EQNNe}
  with the numerical solution of the high-density asymptotic
  approximation derived from using Eqs.~\ref{SILargeLimit} in
  Eqs.~\ref{EQN5}. We plot the total susceptible and infected
  populations, $\Xs+\Xsi+2\Xss$ and $\Xi+\Xsi+2\Xii$, derived from
  Eqs.~\ref{EQNNa}-\ref{EQNNe} (solid black) versus $\NS^{(0)}$ and
  $\NI^{(0)}$ from Eqs.~\ref{SILargeLimit} and \ref{EQN5} (dashed blue
  and dashed red) as functions of $\ln t$. (a) The parameters used are
  $\bar{a}=\bar{d}=1$, $\ve = 0.3$, $\mu_{\rm s}=\mu_{\rm ss}=\mu_{\rm
    si}=0$, $\mu_{\rm i}=\mu_{\rm ii}=\mu_{\rm is}=0.01$, $\Pi=2$, and
  $\beta=0.5$, with initial conditions $\Xs^{(0)}(0) = 100$ and
  $\Xi^{(0)}(0)=10$.  (b) Using the same parameters and initial
  conditions, but with $\ve=0.0003$.  In both plots, the decreasing
  and increasing curves indicate $\NS^{(0)}(t)$ and $\NI^{(0)}(t)$,
  respectively. The asymptotic approximations are quite accurate even
  for $\ve=0.3$.}
\label{FIG1}
\end{center}
\end{figure*}
In Figs.~\ref{FIG1} we compare the exact solutions of $\NS(t)$ and
$\NI(t)$ from Eqs.~\ref{EQNNa}-\ref{EQNNe} to $\NS^{(0)}(t)$ and $\NI^{(0)}(t)$
derived from solving Eqs.~\ref{EQN5} using
Eqs.~\ref{SILargeLimit}. The agreement is excellent for all times.

\subsubsection{Equal association rates and equal dissociation rates}
A further simplification can be made in the special case in which both
the pairing rates and unpairing rates for all possible pairings are
equal.  This implies that the dissociation coefficients for each of
the pairings are the same $d_{\rm ss}=d_{\rm ii}=d_{\rm si}$.  For
association, there are three possible pairings:
susceptible-susceptible, infected-infected, and susceptible-infected.
A pair with one infected and one susceptible can combinatorially arise
in two ways so $a_{\rm si}=2a_{\rm ss}=2a_{\rm ii}$. Thus,
$\kappa_{\rm si}=2\kappa_{\rm ss}=2\kappa_{\rm ii}\equiv \kappa$ and $K=1$. The
physical solution to Eqs.~\ref{TotalDensitiesLeadingOrdera} and
\ref{TotalDensitiesLeadingOrderb} then reduces to
%

\begin{align}
\begin{split}
  F_{\rm S}(\NS^{(0)},\NI^{(0)}) &= \frac{\NS^{(0)}}{4\kappa\left(\NS^{(0)}+\NI^{(0)}\right)}
\left(\sqrt{8\kappa\left(\NI^{(0)}+\NS^{(0)}\right)+1}-1\right),
\\
F_{\rm I}(\NS^{(0)},\NI^{(0)}) &=\frac{\NI^{(0)}}{4\kappa\left(\NS^{(0)}+\NI^{(0)}\right)}
\left(\sqrt{8\kappa\left(\NI^{(0)}+\NS^{(0)}\right)+1}-1\right).
\label{SIK1}
\end{split}
\end{align}

Using these expressions, Eqs.~\ref{EQN5} in the $N_{\rm S}^{(0)}+N_{\rm I}^{(0)}\gg
1$ limit simplify to

\begin{align}
\begin{split}
  \displaystyle  {\dd \NS^{(0)}\over \dd t} & = \Pi -\mu_{\rm s}\NS^{(0)}  
+ (\mu_{\rm s}-\mu_{\rm ss})\frac{\NS^{(0)2}}{\NS^{(0)}+\NI^{(0)}}  +
{(\mu_{\rm s}-\mu_{\rm si}-\beta)\over 2}\frac{\NS^{(0)}\NI^{(0)}}{\NS^{(0)}+\NI^{(0)}}\\
\displaystyle  {\dd \NI^{(0)} \over \dd t} & = -\mu_{\rm i} \NI^{(0)}+(\mu_{\rm i} - \mu_{\rm ii})
\frac{\NI^{(0)2}}{\NS^{(0)}+\NI^{(0)}}  +
{(\mu_{\rm i} -\mu_{\rm is}+\beta)\over 2}\frac{\NS^{(0)}\NI^{(0)}}{\NS^{(0)}+\NI^{(0)}},
\label{SISEqualLargeLimit}
\end{split}
\end{align}
which is similar to a frequency-dependent model with effective
transmission rate $B_{\rm f}=\beta/2$. Thus, we have found a specific limit
where pairing and unpairing dynamics within a mass-action model
reduces it to an effective frequency-dependent model.


\subsection{Fast Association Limit}
We now consider a different limit in which we relax the fast
dissociation constraint and assume only the association rates are
significantly larger than all other (dissociation, death, and
infection) rates.  Upon defining $a_{\rm ss}=\bar{a}_{\rm ss}/\ve$,
$a_{\rm si}=\bar{a}_{\rm si}/\ve$ and $a_{\rm ii}=\bar{a}_{\rm
  ii}/\ve$, with $\ve \to 0^{+}$, Eqs.~\ref{EQNNa}-\ref{EQNNe} become
\begin{align}
\begin{split}
\displaystyle  {\dd \Xs \over \dd t}   &= \Pi -\mu_{\rm s}\Xs 
- 2\frac{\bar{a}_{\rm ss}}{\ve}\Xs^2 -
\frac{\bar{a}_{\rm si}}{\ve}\Xs\Xi + 2(\mu_{\rm ss}+d_{\rm ss})\Xss +
(\mu_{\rm is}+d_{\rm si})\Xsi \\
\displaystyle  {\dd \Xi \over \dd t}   &= -\mu_{\rm i}\Xi -2\frac{\bar{a}_{\rm ii}}{\ve}
\Xi^2 - \frac{\bar{a}_{\rm si}}{\ve}\Xs\Xi + 2(\mu_{\rm ii}+d_{\rm ii})\Xii  +
(\mu_{\rm si}+d_{\rm si})\Xsi \\
\displaystyle  {\dd \Xss \over \dd t}   & = -\left(2\mu_{\rm ss}+d_{\rm ss}\right)\Xss
+\frac{\bar{a}_{\rm ss}}{\ve}\Xs^2 \label{FastAssoc} \\
\displaystyle  {\dd \Xsi  \over \dd t}   &= -\left(\mu_{\rm is}+\mu_{\rm si} +d_{\rm si}
 +\beta\right)\Xsi +\frac{\bar{a}_{\rm si}}{\ve}\Xs\Xi \\
\displaystyle  {\dd \Xii   \over \dd t}   &=  -\left(2\mu_{\rm ii} + d_{\rm ii}\right)
\Xii  +\beta \Xsi+\frac{\bar{a}_{\rm ii}}{\ve}\Xi^2.
\end{split}
\end{align}
Substituting the expansion
\begin{align}
\begin{split}
\Xs &= \Xs^{(0)} + \ve^{1/2} \Xs^{(1)}+ \cdots\\
\Xi &= \Xi^{(0)} + \ve^{1/2}\Xi^{(1)}+ \cdots\\
\Xss &= \Xss^{(0)} + \ve^{1/2} \Xss^{(1)} +\cdots\\
\Xsi &= \Xsi^{(0)} + \ve^{1/2} \Xsi^{(1)}+ \cdots\\
\Xii &= \Xii^{(0)} + \ve^{1/2} \Xii^{(1)} +\cdots
\label{FAExpansion}
\end{split}
\end{align}
into Eqs.~\ref{FastAssoc} and retaining only terms of size
$O\left(\ve^{-1}\right)$, we find $\Xs^{(0)} = \Xi^{(0)}= 0$.  Next,
collecting terms of size $O\left(1\right)$ we obtain

\begin{subequations}
\begin{align}
0   &= \Pi  - 2\bar{a}_{\rm ss}\Xs^{(1)2} -
\bar{a}_{\rm si}\Xs^{(1)}\Xi^{(1)} + 2(\mu_{\rm ss}+d_{\rm ss})\Xss^{(0)} +
(\mu_{\rm is}+d_{\rm si})\Xsi^{(0)} \label{FAAsympta} \\
0   &= -2\bar{a}_{\rm ii}\Xi^{(1)2} -
\bar{a}_{\rm si}\Xs^{(1)}\Xi^{(1)}+ 2(\mu_{\rm ii}+d_{\rm ii})\Xii^{(0)} +
(\mu_{\rm si}+d_{\rm si})\Xsi^{(0)} \label{FAAsymptb} \\
\displaystyle  {\dd \Xss^{(0)} \over \dd t}   & = 
-\left(2\mu_{\rm ss}+d_{\rm ss}\right)\Xss^{(0)} +\bar{a}_{\rm ss}\Xs^{(1)2}
\label{FAAsymptc} \\
\displaystyle  {\dd \Xsi^{(0)}  \over \dd t}   &= -\left(\mu_{\rm is}+\mu_{\rm si} 
+d_{\rm si} +\beta\right)\Xsi^{(0)} +
\bar{a}_{\rm si}\Xs^{(1)}\Xi^{(1)}\label{FAAsymptd} \\ 
\displaystyle  {\dd \Xii^{(0)}   \over \dd t}   &=  
-\left(2\mu_{\rm ii} + d_{\rm ii}\right)\Xii^{(0)}  +\beta \Xsi^{(0)}+\bar{a}_{\rm ii}\Xi^{(1)2}
\label{FAAsympte}
\end{align}
\end{subequations}


Solving Eqs.~\ref{FAAsympta} and \ref{FAAsymptb}, we find 
\begin{align}
\begin{split}
\Xs^{(1)} &= \sqrt{\frac{P_{\rm i}+(2f-1)P_{\rm s} - 
\sqrt{\left(P_{\rm i}-P_{\rm s}\right)^2+4f P_{\rm i} P_{\rm s}}}{4\bar{a}_{\rm ss}(f-1)}}
\\
\Xi^{(1)} &= \sqrt{\frac{P_{\rm s}+(2f-1)P_{\rm i} - 
\sqrt{\left(P_{\rm i}-P_{\rm s}\right)^2+4f P_{\rm i} P_{\rm s}} }{4\bar{a}_{\rm ii}(f-1)}},
\label{SIALarge}
\end{split}
\end{align}
where $f\equiv 4\bar{a}_{\rm ss}\bar{a}_{\rm ii}/\bar{a}_{\rm si}^2$
and

\begin{align}
\begin{split}
P_{\rm s} &= 2(\mu_{\rm ss}+d_{\rm ss})\Xss^{(0)}+(\mu_{\rm is}+d_{\rm si})\Xsi^{(0)}+\Pi\\
P_{\rm i} &= 2(\mu_{\rm ii}+d_{\rm ii})\Xii^{(0)}+ (\mu_{\rm si}+d_{\rm si})\Xsi^{(0)}.
\label{DefAlpha}
\end{split}
\end{align}
Thus, to lowest order in the fast association limit, the infected
population is $\NI^{(0)} \approx 2\Xii^{(0)}+\Xsi^{(0)}$.  In what
follows, it will be useful to define the susceptibles who are in
susceptible-infected pairs, $N_{\rm E}^{(0)}\equiv \Xsi^{(0)}$, as an
``exposed'' population. Analogously, the ``unexposed'' susceptible
population \textit{not} in mixed pairs is dominated by
susceptible-susceptible pairs and is $\NS^{(0)} \approx 2\Xss^{(0)}$.

%

Rewriting Eqs.~\ref{FAAsymptc}-\ref{FAAsympte}
using Eqs.~\ref{SIALarge}, we find

\begin{align}
\begin{split}
\displaystyle  {\dd \NS^{(0)} \over \dd t}   & = 
-\left(2\mu_{\rm ss}+d_{\rm ss}\right)\NS^{(0)}                      +
\frac{P_{\rm i}+(2f-1)P_{\rm s} - \sqrt{\left(P_{\rm i}-P_{\rm s}\right)^2
    +4f P_{\rm i} P_{\rm s}}}{2(f-1)}
\\
\displaystyle  {\dd N_{\rm E}^{(0)}  \over \dd t}   &= 
-\left(\mu_{\rm is}+\mu_{\rm si} +d_{\rm si} +\beta\right)N_{\rm E}^{(0)} +
\frac{ \sqrt{\left(P_{\rm i}-P_{\rm s}\right)^2+4f P_{\rm s} P_{\rm i}}
  -\left(P_{\rm i}+P_{\rm s}\right)}{2(f-1)}
\\
\displaystyle  {\dd \NI^{(0)}  \over \dd t}   &=
-\mu_{\rm ii} \NI^{(0)}
+(\mu_{\rm ii}-\mu_{\rm is}+\beta)N_{\rm E}^{(0)},
\label{FAAsympt2}
\end{split}
\end{align}
where $P_{\rm s}$ and $P_{\rm i}$ can also be expressed as 
\begin{align}
\begin{split}
  P_{\rm s} &= (\mu_{\rm ss}+d_{\rm ss})N_{\rm S}^{(0)} + (\mu_{\rm is}+d_{\rm si})
  N_{\rm E}^{(0)} + \Pi
\\
P_{\rm i} &= (\mu_{\rm ii}+d_{\rm ii})N_{\rm I}^{(0)} +  (\mu_{\rm si}-\mu_{\rm ii}
+d_{\rm si}-d_{\rm ii})N_{\rm E}^{(0)}.
\label{DefAlpha2}
\end{split}
\end{align}
Eqs.~\ref{FAAsympt2} and \ref{DefAlpha2} constitute a self-contained
system of equations for the three subpopulations $\NS^{(0)}(t), N_{\rm
  E}^{(0)}(t)$, and $\NI^{(0)}(t)$.

\begin{figure*}[htb]
\begin{center}
\includegraphics[width=4.4in]{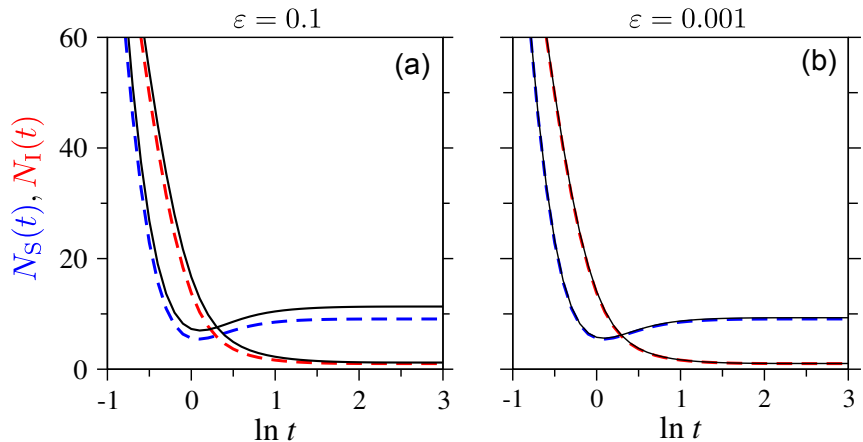}
\caption{{\bf Fast association limit:} Comparison of exact numerical
  solution of Eqs.~\ref{EQNNa}-\ref{EQNNe} to the numerical solution
  of the high association rate approximation (Eqs.~\ref{FAAlt}). The
  exact solutions of $\Xs+\Xsi+2\Xss$ and $\Xi+\Xsi+2\Xii$ from
  Eqs.~\ref{EQNNa}-\ref{EQNNe} are compared with the corresponding
  quantities $\NS +N_{\rm E}$ and $\NI$ found from numerically
  integrating Eqs.~\ref{FAAlt}.
%
%
  (a) The parameters used are $\bar{a}=1, d=1$, $\mu_{\rm s}=0.01,
  \mu_{\rm i}=\mu_{\rm is}=\mu_{\rm si}=0.05$, $\mu_{\rm ii}=\mu_{\rm
    ss}=5$, $\Pi=50$, and $\beta=100$, with initial conditions $\Xs(0)
  = \Xi(0)=0, \Xss(0) = 500, \Xsi(0)=100$, and $N_{\rm ii}=10$.  The
  corresponding initial conditions for Eqs.~\ref{FAAlt} are
  $\NS(0)=1000$, $N_{\rm E}(0)=100$, and $\NI(0)=120$.  The
  approximation is highly accurate even for $\ve = 0.1$. In (b), we
  used the same parameters but set $\ve=0.001$. The agreement is also
  quite good and improves as $\ve$ decreases.}
\label{FIG2}
\end{center}
\end{figure*}

An alternative formulation is to group all susceptibles and write

\begin{align}
\begin{split}
\displaystyle  {\dd \over \dd t}\left(\NS^{(0)}+N_{\rm E}^{(0)} \right)  & 
= \Pi -\mu_{\rm ss} \left(\NS^{(0)}+N_{\rm E}^{(0)}\right)      +
(\mu_{\rm ss}-\mu_{\rm si}-\beta)N_{\rm E}^{(0)} \\
\displaystyle  {\dd N_{\rm E}^{(0)}  \over \dd t}   &
= -\left(\mu_{\rm is}+\mu_{\rm si} +d_{\rm si} +\beta\right)N_{\rm E}^{(0)} +
\frac{\sqrt{\left(P_{\rm i}-P_{\rm s}\right)^2+4 f P_{\rm s} P_{\rm i}} -\left(P_{\rm i}+P_{\rm s}\right)}{2(f-1)}
\\
\displaystyle  {\dd \NI^{(0)}  \over \dd t}   &=
-\mu_{\rm ii}\NI^{(0)}
+(\mu_{\rm ii}-\mu_{\rm is}+\beta)N_{\rm E}^{(0)}.
\label{FAAlt}
\end{split}
\end{align}
In the case $\bar{a}_{\rm si}^2 \to 4\bar{a}_{\rm ss}\bar{a}_{\rm ii}$
($f \to 1$), we apply L'Hopital's rule on Eqs.~\ref{FAAsympt2} to
further simplify it to

\begin{align}
\begin{split}
\displaystyle  {\dd \NS^{(0)} \over \dd t}   & = -\left(2\mu_{\rm ss}+d_{\rm ss}\right)
\NS^{(0)} + \frac{P_{\rm s}^2}{P_{\rm s}+P_{\rm i}}\\
\displaystyle  {\dd N_{\rm E}^{(0)}  \over \dd t}   &= -\left(\mu_{\rm is}+\mu_{\rm si} 
+d_{\rm si} +\beta\right)N_{\rm E}^{(0)} +
\frac{P_{\rm s} P_{\rm i}}{P_{\rm s}+P_{\rm i}}
\\
\displaystyle  {\dd \NI^{(0)}  \over \dd t}   &=
-\mu_{\rm ii} \NI^{(0)}
+(\mu_{\rm ii}-\mu_{\rm is}+\beta)N_{\rm E}^{(0)},
\label{FACompSEI_A1}
\end{split}
\end{align}
which is reminiscent of simple SEI-type models
\citep{SEIR1,HETHCOTE2000,SEIR0}.  A comparison between $\NS$ and $\NI$
derived from the exact Eqs.~\ref{EQNNa}-\ref{EQNNe} and those derived
from solving Eqs.~\ref{FACompSEI_A1} is given in Figs.~\ref{FIG2}.
The approximations are accurate for all valid parameter regimes across
all times.

\section{Summary and Conclusions}

We have revisited the canonical mass-action susceptible-infected
disease transmission models and systematically incorporated pairing
dynamics.  The purpose is to rigorously find uniformly valid effective
equations from mass-action models with pairing and unpairing steps.
After nondimesionalization of the five fundamental mass-action
equations, we find parameter regimes that allow us to develop
uniformly valid approximations to the total infectious and susceptible
populations.  Our results were compared with lower-dimensional mass-action and
frequency-dependent models \textit{without} pairing.

First, in the fast transmission and pair dissociation limit, we found
that the mass-action pairing model reduces to a standard mass-action
susceptible-infected (SI) model (Eqs.~\ref{EQNEFF}) without pairing,
but with an effective disease transmission rate given by
Eq.~\ref{BETAEFF}.

Next, if pair formation \textit{and} break-up are assumed fast, we
found effective equations for the total susceptible and infected
populations.  Although the resulting two ODEs can be unwieldy, this
system differs fundamentally from the basic SI model. However, if
death rates do not depend on the pairing status, we show that, in the
low-density limit, the simple mass-action response is recovered
(Eq.~\ref{SISSmallLimit}).  In this low-density limit, the pairing
dynamics do not affect the leading-order form of the functional
response.
%
%
However, in the high-density limit, a frequency-dependent response is
recovered (Eq.~\ref{SISEqualLargeLimit}) if the association and
dissociation rates are the same for each of the three different types
of pairs.  Under these assumptions, we showed that, for finite
densities, a Holling's Type II response does not arise. Nevertheless,
we derived a simple functional response that contains the same number
of parameters as a model using Holling's type II response, but with a
clear mathematical justification.

Finally, we relaxed the fast dissociation constraint and assumed that
only the association rates are large. In this case, we could only
reduce the five-dimensional system of mass-action equations to a
three-dimensional system that includes susceptible, infecteds, and an
exposed population describing susceptible in susceptible-infected
pairs (Eqs.~\ref{FAAsympt2} or \ref{FACompSEI_A1}). These equations
share features with the canonical SEI-type models
\citep{ANDERSON_MAY1991,HETHCOTE2000,SEIR1}.

Although the two- or three-dimensional system of equations we derived
are typically more complicated in form, our formulae allow one to
incorporate the effects of pair formation and dissociation in a
self-consistent, uniformly valid way in the limits described.  We have
also numerically compared our solutions with those from the full
five-dimensional mass-action system and found excellent agreement in
the limits analyzed (Figs.~\ref{FIG1} and \ref{FIG2}).

Our asymptotic analysis can be, in principle, straightforwardly
extended to richer disease models including SIS, SIR, and models with
birth processes. It would be interesting to apply similar asymptotic
approaches to analyze disease dynamics occurring under group
interactions or household structures
\citep{SANDER2004,HOUSEHOLD2007,HOUSEHOLD2008,NETWORK} or with aging
\citep{AGINGJSP}.

Mass-action chemical reaction models in which an enzyme and substrate
must first associate before a reaction can occur have also been
treated using related asymptotic analyses. A classic example is
Michaelis-Menten kinetics in which an inner and outer solution are
pieced together to describe substrate and product concentrations at
short and long times \citep{KEENER}.  In our problem, we have only
considered the ``outer'' solutions, yet for all cases studied, our
lowest order approximations are valid at all times. A more detailed
and rigorous examination of bimolecular interactions in mass-action
chemical kinetics in certain reaction rate limits may provide new
mathematical insights on approximating complex reaction networks.

\vspace{3mm}

\section*{Acknowledgements}
JW was supported by the Research Grants Council of Hong Kong Special
Administrative Region through project CityU11306115. TC was supported
by the NSF through grant DMS-1814364, the Army Research Office through
grant W911NF-18-1-0345,
%
%
and the Beijing Computational Science Research Center. The authors
thank X. Cheng for useful feedback on the manuscript.

\bibliographystyle{spbasic}
\bibliography{refs}

\end{document}